# Interference-type plasmonic polarizers and generalized law of Malus

Cheng-ping Huang*, Yu-lin Wang, and Yong Zhang

*Department of Physics, Nanjing Tech University, Nanjing 211816, China*

*Email: cphuang@njtech.edu.cn

The conventional linear polarizer only allows the electric component parallel to the polarizer axis to pass through whereas prohibits the vertical component. We propose that a specially-designed single-layer plasmonic polarizer can couple both parallel and vertical electric components to the transmission, thus breaking the classical law of Malus. A variety of anomalous polarization effects, such as the asymmetric polarization-angle dependence, enhanced polarization filtering with wide polarization angle, and tunable polarization rotation from $0^o$ to $90^o$, can be resulted. To understand the effects, the generalized law of Malus, originating from the superposition principle, has been presented and analyzed. This provides a basis for studying the interference-type plasmonic polarizers, where the interference effect and polarization effect are combined together. The difference between the plasmonic and conventional polarizers is of both fundamental and practical interests.

## 1. Introduction

Recently, quasi-two-dimensional subwavelength plasmonic nanostructures or metasurfaces, such as the periodic arrays of metallic nanoantennas or metal films milled with small apertures, have received much research interest in the scientific community [1-3]. The reasons are as follows: the plasmonic nano-structures are artificially constructed and only limited by the imagination, physics rich and multifunctional for controlling light (e.g., the amplitude, polarization, phase, and propagation direction etc.), and size compact and suitable for photonic integration or of many other potential applications. For example, with the gradient metasurfaces composed of nanoantennas of varying parameters, anomalous light reflection, refraction, and diffraction have been studied [4-7]. Moreover, the gradient metasurfaces enable the construction of planar metalenses with the subwavelength resolution [8-10].

The metasurfaces also play a crucial role in manipulating the polarization of light [11-14]. It is well known that, besides the enhanced optical transmission effect, a metal film milled with one-dimensional slits or two-dimensional rectangular holes behaves as a nanoscale plasmonic polarizer [15, 16]. The effect is associated with the excitation and propagation of waveguide modes in the subwavelength apertures. Conventionally, the electric component parallel to the polarizer axis is allowed to pass through the polarizer whereas the vertical component is prohibited. This is the physical origin of the classical Malus' law. Similar effect can also be found with the single-layer plasmonic polarizers [15, 16]. Recent theoretical and experimental studies suggested that the transmission response of multiple plasmonic polarizers does not follow the prediction of classical Malus' law [17-19]. One example is that, with the use of three cascaded plasmonic polarizers, wide-band and highly efficient (~80%) polarization rotation can be achieved [19-22]. Another example is that two orthogonal plasmonic polarizers, which will usually forbid the transmission of light, can induce high transparency and 90-degree polarization rotation [17, 23]. These effects stem from the near-field coupling and/or multiple reflections in the plasmonic systems.

Compared with the multi-layer systems, the single-layer plasmonic structure owns some advantages such as the convenience of fabrication and device integration. One may ask: may we construct a single-layer plasmonic polarizer that couples both the parallel and vertical electric components of incident light to the transmitted polarization? And, may we use this polarizer to transform the natural light into linear polarization with the efficiency larger than 50% or even up to 100%? These questions are of both theoretical and practical importance. In this paper, we propose that a specially-designed single-layer plasmonic polarizer, consisting of a metal film milled with rectangular holes and grooves which are orthogonal to each other, may realize the first function and thus break the classical law of Malus. With the polarizer, some abnormal polarization properties such as the asymmetric polarization-angle dependence, efficient polarization generation with wide polarization angle, and tunable polarization rotation from $0^o$ to $90^o$ (by rotating the polarizer mechanically), etc. can be achieved. Moreover, based on the superposition principle, the generalized law of Malus describing the relationship between the transmission and polarization angle has been presented and analyzed. We also prove theoretically that the conversion efficiency from natural light to linear polarization will not exceed 50% with such a polarizer.

We emphasize that the polarizer called here "plasmonic" is mainly due to the fact that the polarizer is composed of plasmonic materials rather than other media. Actually, even if the metal is a perfect electric conductor and the plasmonic effect is not present, similar result can still be obtained with the structure (not shown here). Nonetheless, for the real metal, the waveguide mode in the rectangular apertures will show the plasmonic character (because of the coupling to the free electrons, the waveguide mode is bounded near the hole walls [24, 25]). This modifies the effective index of waveguide and shifts the peak position to the longer wavelength.

In the following, the paper is divided into four sections. In Sec. 2, the structure design of the plasmonic polarizer and the method of investigation were illustrated. In Sec. 3, numerical simulations were implemented and the anomalous polarization effects of the polarizer were presented. The generalized law of Malus was presented and analyzed in Sec. 4. And a short summary was provided in Sec. 5.

## 2. Structure and method

Figure 1 shows the schematic view of the single-layer plasmonic polarizer, which is composed of a metal (silver) film perforated with a square array of specially-designed subwavelength apertures. For simplicity, the perforated metal film is assumed to be freestanding (the fabrication and optical properties of free-standing metal films have been reported experimentally [26, 27]; the qualitative conclusion of this paper can be extended to the metal film with a substrate). In the metal film, the rectangular holes and grooves are perforated orthogonally, which form the Γ-shaped crossing in the $xy$ plane. The period of the square lattice is $p$, the arm length and width of the combined holes and grooves are $l$ and $w$, respectively (see inset in the lower right corner of Fig. 1). The thickness of metal film is $t$ and the depth of rectangular groove is $h$. A linearly polarized light is incident normally upon and transmits through the metal film (the propagation direction is along the $+z$ axis). The polarization angle of incident light is assumed to be $\theta$ with respect to the $y$ axis.

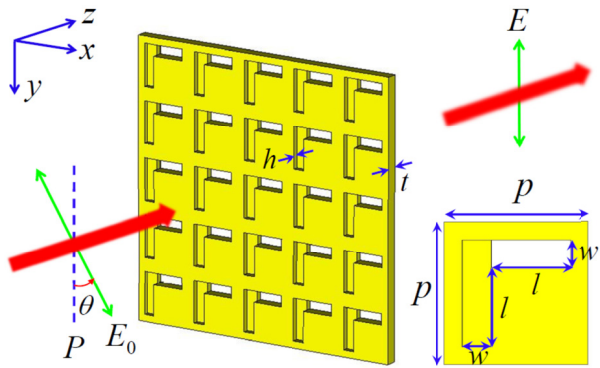

Fig. 1 Schematic view of a single-layer of interference-type plasmonic polarizer. The polarization axis of the polarizer, denoted by P, is along the y direction. The linearly polarized light with an inclined polarization (the polarization angle is θ with respect to the y axis) is incident normally upon and transmits through the polarizer.

In this paper, our interest is focused on the theoretical investigation of the plasmonic polarizer in the optical regime. Without loss of generality, the structural parameters of the system are set as follows throughout the paper. The lattice period is $p$=1000 nm, the arm length and width are $l$=550 nm and $w$=200 nm, the thickness of metal film is $t$=450 nm, and the groove depth is $h$=200 nm (the metal under the bottom of groove is much thicker than the skin depth, thus forbidding the transmission of light through the groove bottom). The variation of structural parameters will not change the resulting conclusion. To study the polarization effect of the structure, numerical simulation based on the finite-difference time-domain (FDTD) method has been implemented, employing the commercial software package FDTD Solutions 6.5 (Lumerical Solutions, Inc., Canada). The periodic boundary conditions in the $xy$ plane and open boundary condition in the $z$ direction have been employed. In the optical frequency range, the dispersion of silver film was described by the Drude model $\varepsilon_m = 1 - \omega_p^2 / \omega(\omega + i\gamma)$ [28], with the plasma frequency of $\omega_p = 1.37 \times 10^{16}$ rad/s and the electron collision frequency of γ=5×10$^{13}$ Hz [to account for the extra electron scattering from the perforated apertures in the silver film, here a larger γ compared with that of Ref. [28] ($\gamma = 3.2 \times 10^{13}$ Hz) is used]. In addition, the electric field of the linearly polarized incident light was set as $E_0$=1 V/m in the simulation.

## 3. Anomalous polarization effects

Figure 2(a) presents the simulated spectra of the field transmission coefficients for $y$- and $x$-polarization of incident light. When the incident light is polarized along the $y$-axis ($E_0=E_{0y}$), the field transmission coefficient $t_{yy}=E_{ty}/E_{0y}$ ($E_{ty}$ represents the real amplitude of the $y$ component of the transmitted electric field; see the solid line) exhibits two peaks around the wavelength of 1127 nm ($t_{yy}$=0.716) and 1530 nm ($t_{yy}$=0.645). To understand the mechanism of the two transmission peaks, Figs. 2(c) and 2(d) plot the electric-field distributions of the apertures for the wavelength 1127 nm and 1530 nm, respectively. One can see that the rectangular holes and grooves can be excited by the $y$-polarized incident light simultaneously. Actually, in this composite plasmonic system, the rectangular holes and grooves are coupled together, giving rise to the mixed or hybrid waveguide modes (in the conjunct part with a depth of $h$). With respect to the diagonal line of unit cell [the dash line in Fig. 2(c)], the fields in the holes and grooves show the antisymmetric- or symmetric- like characteristic. Thus, the two transmission peaks are correlated with the mixed antisymmetric and symmetric waveguide modes. One can also see that, compared with Fig. 2(c), the fields in the rectangular groove of Fig. 2(d) are relatively weak. This point can be understood as follows. In the groove, a super-position of forward wave and backward wave reflected at the groove bottom can be resulted. At the longer wavelength [Fig. 2(d)], the groove will be far from the standing wave resonance state, thus yielding a weak field enhancement. In addition, the amplitude of x-component of the transmitted field $E_{tx}$ was simulated. We found that the transmission coefficient $t_{xy}=E_{tx}/E_{0y}$ is very close to zero within the studied wavelength range ($t_{xy}\sim10^{-3}$). The effect was robust regardless of the incident polarization direction. Hence, the single-layer perforated metal film works as a miniature plasmonic polarizer, allowing the polarization along the y-axis to be filtered out (i.e., the polarizer axis is in y direction).

Nonetheless, this is not a usual plasmonic polarizer. When the incident light is polarized perpendicular to the polarizer axis (i.e., along the x-axis with $E_0=E_{0x}$), the field transmission coefficient $t_{xx}=E_{tx}/E_{0x}$ is near zero as expected but $t_{yx}=E_{ty}/E_{0x}$ also presents two transmission peaks [see Fig. 2(a), the dash line]. This point is distinct from the conventional polarizers; usually, the transmission of light will be prohibited when the incident polarization is perpendicular to the polarizer axis. The results indicate that the plasmonic polarizer can couple both the parallel and vertical electric components to the transmission. In addition, due to the special design of structure, the effect also makes a difference from the depolarization phenomenon: In our case, the incident polarized or unpolarized light can be always converted to the transmitted linear polarization with one fixed direction; in the depolarization case, the incident linearly-polarized light will induce the less-polarized (e.g., partially-polarized or unpolarized) transmission with two orthogonal electric components. The positions of two transmission peaks are shifted slightly to the wavelength of 1200 nm ($t_{yx}$=0.747) and 1520 nm ($t_{yx}$=0.243). Figures 2(e) and 2(f) plot the electric-field distributions of the apertures for the two peaks, respectively. We can see that the character of field distributions is similar to that of Figs. 2(c) and 2(d), suggesting that the transmission peaks share a common physical origin. We noticed that the peak of $t_{yx}$ around 1530 nm is much lower than that of $t_{yy}$. The reason is that, for x-polarization, the transmission through the rectangular holes is mediated by the groove-hole coupling effect (rather than induced directly by the incident light). As mentioned above, the groove waveguide mode will be weakly excited at the longer wavelength, thus suppressing the groove-hole coupling and the transmission.

Along with the field transmission coefficients $t_{yy}$ and $t_{yx}$ (which dominate the optical properties of the plasmonic polarizer), the corresponding phase responses were simulated as well and the phase difference between them $\delta=\varphi_{yy}-\varphi_{yx}$ is mapped in Fig. 2(b). Here, $\varphi_{yy}$ and $\varphi_{yx}$ represent the phases of y-component of the transmitted field with the incident polarization along the y- and x-axis, respectively. With the increase of wavelength, the phase difference $\delta$ varies gradually from ~180° to ~0°. The phase evolution characteristic of the far fields is in accordance with the near field distributions in the rectangular holes (as shown in Figs. 2c-2f), which show the transition from antisymmetric to symmetric mixed waveguide mode. The phase difference will be used in Sec. 4 to develop a quantitative analysis. We mention here that the transmission coefficients and phase difference are dependent on the detailed structural parameters (such as the groove depth, film thickness, etc.), which offers multiple degrees of freedom for optimizing the performance of the polarizer.

Then how is the polarization effect when the incident light is polarized with an arbitrary polarization angle? To answer this question, the power transmission efficiency for inclined incident polarization were simulated, taking $\theta=\pm45°$ as two examples (the power transmission efficiency of the polarizer is defined as T=$T_y$=($E_{ty}/E_0$)$^2$; note that $T_x$=($E_{tx}/E_0$)$^2$ is very small and not shown here and hereafter). The results are presented in Fig. 3(a, b) by the solid lines. A significant characteristic of the plas-

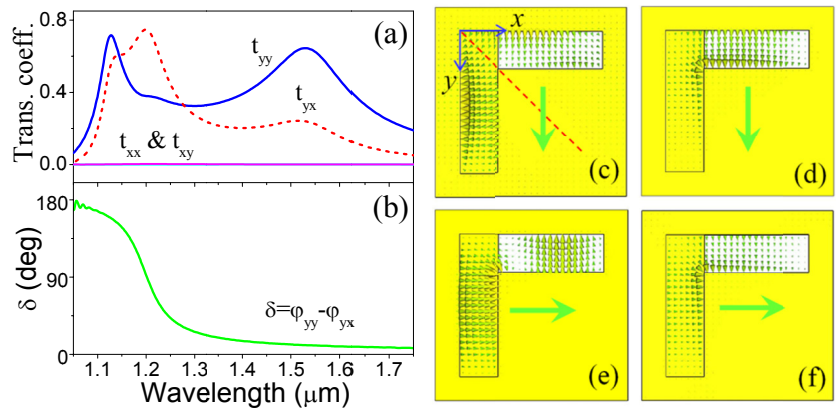

Fig. 2 (a) Electric-field transmission coefficients for y- and x-polarized incident light. (b) Phase difference δ as a function of wavelength. (c, d) Electric-field distributions of the apertures with the excitation of y-polarized light: (c) λ=1127 nm and (d) λ=1530 nm; (e, f) Electric-field distributions with the excitation of x-polarized light: (e) λ=1200 nm and (f) λ=1520 nm. The thick (green) arrows indicate the incident electric field.

monic polarizer is that the transmission spectra are strongly asymmetric for ±$\theta$, which also violates the classical law of Malus. When $\theta=45^o$ [see Fig. 3(a)], two peaks appear at the wavelength 1215 nm and 1524 nm, with the transmission efficiency close to 40%. On the other hand, when $\theta=-45^o$ [see Fig. 3(b)], two transmission peaks can also be achieved: the peak at 1536 nm has a lower efficiency of 8.5%; the peak at 1132 nm owns a maximal efficiency up to 80%, which is significantly larger than that obtained with the y- or x-polarization ($t_{yy}^2 \leq 51\%$ and $t_{yx}^2 \leq 56\%$).

To further investigate the dependence of power transmission efficiency on the polarization angle of the incident light, numerical simulations have been implemented by setting the working wavelength as 1132 nm and 1524 nm (around the two transmission peaks). The results are shown in Fig. 3(c) (1132 nm) and 3(d) (1524 nm) with the solid lines, respectively. For $\theta$ ranging from $-90^o$ to $90^o$, the asymmetric character can be seen clearly. For the wavelength 1132 nm [see Fig. 3(c)], the power transmission efficiency is higher than 34% when $\theta$ varies from $0^o$ to $-90^o$ ($T>47\%$ when $0^o \geq \theta \geq -80^o$) and a maximal efficiency of 80.4% is obtained when $\theta=-40^o$. The wide-angle and efficient polarization filtering effect is not available in the conventional polarizers. It is worthy of noticing that, by fixing the polarization direction of incident light (along the original y axis) and rotating the plasmonic polarizer counterclockwise, the polarization angle $\theta$ can also be altered from $0^o$ to $-90^o$. Correspondingly, a counterclockwise, continuous, and efficient polarization rotation of the transmitted field from $0^o$ to $90^o$ can be obtained.

Nonetheless, when the polarization angle $\theta$ varies from $0^o$ to $90^o$, the power transmission efficiency will be degenerated significantly, as shown in the right part of Fig. 3(c). A minimum of transmission ~3.3% appears with the polarization angle $\theta=50^o$. To improve the polarization filtering effect with positive angle $\theta$, one may change the working wavelength from 1132 nm to 1524 nm. Figure 3(d) shows that for the wavelength 1524 nm the transmission efficiency is larger than 27% when $\theta$ varies from $-20^o$ to $60^o$, with the maximal efficiency ~47% locating at $\theta=20^o$. Moreover, the minimum of transmission appears at $\theta=-70^o$ with the efficiency near zero (0.2%). The maxima and minima of transmission are correlated with the near fields of the rectangular holes. Figures 3(e, f) and 3(g, h) plotted the electric-field distributions associated with the transmission maxima or minima of Fig. 3(c) and 3(d), respectively. The results suggest that the hole waveguide mode is strongly boosted at the transmission maxima and suppressed at the minima. We noticed that the cutoff wavelength of the plasmonic rectangular hole, $\lambda_c = 2(l+2\delta_m)\sqrt{1+2\delta_m/w}$ [12, 24], is around 1312 nm ($\delta_m$=22 nm is the skin depth of metal). Consequently, the hole waveguide mode shows propagating character at 1132 nm [Fig. 3(e, f)] and evanescent character at 1524 nm [Fig. 3(g, h)].

## 4. Generalized law of Malus

The anomalous polarization phenomena can be attributed to the interference effect between two transmission channels. When the light with an inclined polarization is incident normally upon the plasmonic polarizer, the electric field can be decomposed into two orthogonal components: $E_{0y} = E_0 \cos\theta$ along the y-axis and $E_{0x} = E_0 \sin\theta$ along the x-axis. The y electric component can induce a transmission through the rectangular holes directly (the first channel); meanwhile, due to the coupling between the grooves and holes, the x electric component can induce a transmission through the rectangular holes indirectly (the second channel). By using the Jones matrix, the complex amplitudes of the transmitted electric fields can be written as

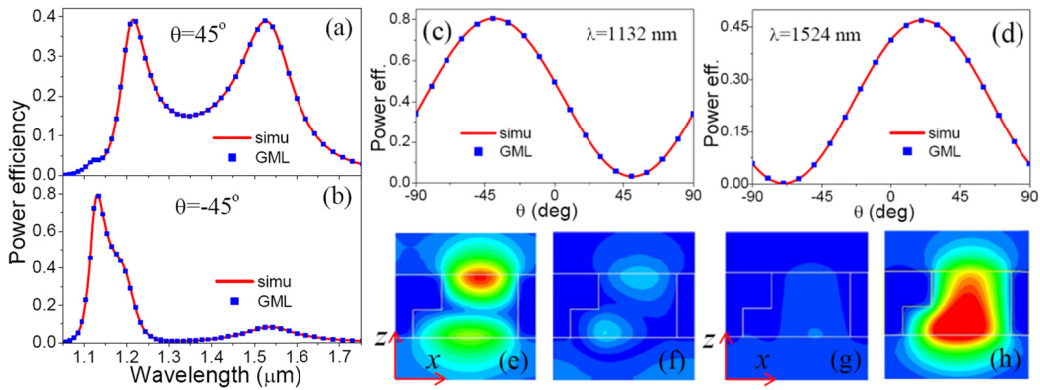

Fig. 3 (a, b) Simulated (the solid lines) power transmission efficiency as a function of wavelength. The polarization angle of the incident light is: (a) $\theta=45^o$ and (b) $\theta=-45^o$. (c, d) Simulated (the solid lines) power transmission efficiency as a function of polarization angle. The wavelength of incident light is: (c) $\lambda=1132$ nm and (d) $\lambda=1524$ nm. The squares represent the results obtained with the generalized Malus' law (GML). The electric-field ($E_y$) distributions (on the xz plane) for the transmission maximum or minimum in (c) and (d) are plotted with (e, f) and (g, h), respectively.

$$\begin{pmatrix} \varepsilon_x \\ \varepsilon_y \end{pmatrix} = \begin{pmatrix} 0 & 0 \\ t_{yx}e^{i\varphi_{yx}} & t_{yy}e^{i\varphi_{yy}} \end{pmatrix} \begin{pmatrix} E_{0x} \\ E_{0y} \end{pmatrix}. \quad (1)$$

Here, $t_{yy}$ ($t_{yx}$) and $\varphi_{yy}$ ($\varphi_{yx}$) are the transmission coefficient and phase as mentioned previously (see Fig. 2). Then, the transmitted field of $x$ component is $\varepsilon_x = 0$ ($t_{xx}$ and $t_{xy}$ are zero according to Fig. 2) and the transmitted field of $y$ component becomes

$$\varepsilon_y(\theta) = E_0 \cos\theta \cdot t_{yy} e^{i\varphi_{yy}} + E_0 \sin\theta \cdot t_{yx} e^{i\varphi_{yx}}. \quad (2)$$

The two terms on the right-hand side of Eq. (2) correspond to the transmission of $y$ and $x$ electric components of incident wave, respectively. Equation (2) is nothing but the well-known superposition principle: The total transmission response of a plasmonic metasurface induced by an inclined incident polarization is the sum of the individual responses that would have been generated by the two orthogonal electric components separately. The superposition of the two individual transmission responses may give rise to a strong constructive or destructive interference effect in certain conditions. For this reason, the plasmonic polarizer investigated in this paper was termed as the interference-type plasmonic polarizer.

With the Eq. (2), the power transmission efficiency of the interference-type plasmonic polarizer, $T(\theta) = |\varepsilon_y(\theta)/E_0|^2$, is expressed as

$$T(\theta) = t_{yy}^2 \cos^2\theta + t_{yx}^2 \sin^2\theta + t_{yy}t_{yx}\cos\delta \sin 2\theta, \quad (3)$$

where $\delta = \varphi_{yy} - \varphi_{yx}$ is the phase difference. Equation (3) establishes a direct relationship between the power transmission efficiency and the polarization angle $\theta$. The transmission efficiency is also a function of wavelength, as both the transmission coefficients and phase difference vary with the wavelength. It should be emphasized that the last term in the Eq. (3) represents the interference contribution, which depends on not only the phase difference $\delta$ but also the polarization angle $\theta$. As the interference term is an odd function of $\theta$, the spectral asymmetry for $\pm\theta$ will be resulted. In addition, when the transmission coefficient $t_{yx}=0$, the Eq. (3) will be degenerated into $T(\theta) = t_{yy}^2 \cos^2\theta$, which corresponds to the classical law of Malus (but modified by the transmission of metasurface). As an extension, the Eq. (3) may be called the generalized Malus' law (GML). In Fig. 3, the transmission of polarizer as a function of wavelength (a, b) or polarization angle (c, d) was also calculated with the Eq. (3) and plotted with the squares (employing the transmission coefficients and phase difference in Fig. 2). As expected, a consistency between GML (the squares) and FDTD simulation (the lines) can be seen.

With the GML, the polarization angle corresponding to the transmission maximum or minimum can be determined simply, satisfying the following equation:

$$\tan 2\theta_m = \frac{2t_{yy}t_{yx}}{t_{yy}^2 - t_{yx}^2}\cos\delta. \quad (4)$$

At the wavelength of transmission peaks (such as 1132 nm and 1524 nm), the phase difference $\delta$ is close to 180° or 0°, as can be seen from Fig. 2(b). With the Eq. (4), the angle $\theta$ for the transmission maximum and minimum obeys approximately

$$\tan\theta_{\max} = -\frac{t_{yx}}{t_{yy}}, \quad \tan\theta_{\min} = \frac{t_{yy}}{t_{yx}} \quad (5a)$$

when $\delta \approx 180°$ ($\cos\delta \approx -1$) or

$$\tan\theta_{\max} = \frac{t_{yx}}{t_{yy}}, \quad \tan\theta_{\min} = -\frac{t_{yy}}{t_{yx}} \quad (5b)$$

when $\delta \approx 0°$ ($\cos\delta \approx 1$). Equations 5(a) and 5(b) can give very accurate value of the angle. At the wavelength 1132 nm, for example, $\delta=157°$, $t_{yy}=0.704$, and $t_{yx}=0.584$. With the Eq. (5a), we have $\theta_{\max} = -39.7°$ and $\theta_{\min} = 50.3°$, which agree well with the simulation results -40° and 50° [see Fig. 3(c)]. And, at the wavelength 1524 nm, $\delta \approx 11°$, $t_{yy}=0.643$, and $t_{yx}=0.242$. With the Eq. (5b), $\theta_{\max} = 20.6°$ and $\theta_{\min} = -69.4°$, in good accordance with the simulation results 20° and -70° [see Fig. 3(d)].

At the given wavelength, the maximal and minimal power transmission efficiency for $\theta_{max}$ and $\theta_{min}$ can be obtained with the GML as follows:

$$T(\theta_m) = \left(t_{yy}^2 + t_{yx}^2 \pm \sqrt{t_{yy}^4 + t_{yx}^4 + 2t_{yy}^2 t_{yx}^2 \cos 2\delta}\right)/2. \quad (6)$$

When the phase difference $\delta \approx 180°$ or 0°, the maximal and minimal power efficiency can be obtained approximately as

$$T_{\max} = t_{yy}^2 + t_{yx}^2, \quad T_{\min} = 0. \quad (7)$$

According to Eq. (7), the approximated maximal efficiency at 1132 nm and 1524 nm is 83.7% and 47.2% respectively, which are close to the simulated values of 80.4% and 47% (see Fig. 3). In addition, the simulated minimal efficiency at 1132 nm and 1524 nm is 3.3% and 0.2% respectively (see Fig. 3), which are close to zero.

Due to the energy conservation law, the maximal power transmission efficiency $T_{max}$ (under the polarization angle $\theta_{max}$) will be less than unity. Therefore, an inequality linking $t_{yy}$ and $t_{yx}$, the electric-field transmission coefficients for the $y$- and $x$-polarized incident light, can be resulted:

$$t_{yy}^2 + t_{yx}^2 \le 1. \quad (8)$$

The inequality implies that there may be a trade-off between $t_{yy}$ and $t_{yx}$: the stronger the transmission for the *y* incident polarization, the weaker the transmission for the *x* incident polarization, and vice versa. The trade-off stems from the coupling effect between the rectangular grooves and holes with the reason as follows. For *y* incident polarization, besides the co-polarized transmission and reflection (induced by the rectangular holes), there is another important energy leakage channel: the cross-polarized reflection from the grooves (mediated by the hole-groove coupling). Generally, the stronger the transmission for the *y* polarization, the weaker the energy leakage from the cross-polarized reflection, and the weaker the hole-groove coupling effect. A weaker hole-groove coupling will give rise to a lower transmission for the *x* incident polarization. It should be pointed out that the inequality (8) also sets a limit for the conversion from the natural light to the linearly polarized light. Since both the *x* and *y* components of the incident fields can be coupled into the plasmonic polarizer, one may consider to transform the natural light into pure *y* polarization with high or even perfect efficiency (50~100%). Actually, in the case of natural incident light, the transmission efficiency of the polarizer can be shown as

$$T_{nl} = (t_{yy}^2 + t_{yx}^2)/2. \qquad (9)$$

Because of the inequality (8), the conversion efficiency will not exceed 50%. This point is identical to the conventional polarizers. Physically, for the natural incident light, the two orthogonal electric components are not coherent and the interference effect is not present in the system. Accordingly, the energy is shared by the reflection (the reflection cannot be suppressed because of the absence of interference), lowering the transmission efficiency.

The GML provides us not only a useful guide to understand the underlying physics but also an efficient tool to calculate the properties of polarizer quickly (once the transmission coefficients $t_{yy}$ and $t_{yx}$ and the phase difference $\delta$ are numerically obtained), without resorting completely to the numerical simulations which are time consuming. To obtain the complete polarization properties of the polarizer, the power transmission efficiency as a function of wavelength and polarization angle has been calculated with the GML and the results are mapped in Fig. 4. Besides the spectral asymmetry with respect to $\theta=0^\circ$, two transmission energy bands can be seen clearly. The center of the lower energy band locates around the wavelength 1520 nm (1420~1620 nm) and the polarization angle $20^\circ$ (-30°~70°). The efficiency of this band ranges from ~20% to ~40%. The higher energy band extends from ~1100 nm to ~1250 nm with the efficiency of 20%~80%. The narrow red region around 1130 nm highlights the transmission enhancement (~80%) due to the strong constructive interference. Although the higher energy band is split into two parts by a narrow blue gap (because of the destructive interference effect), the wide-angle operation character of the plasmonic polarizer is clearly manifested.

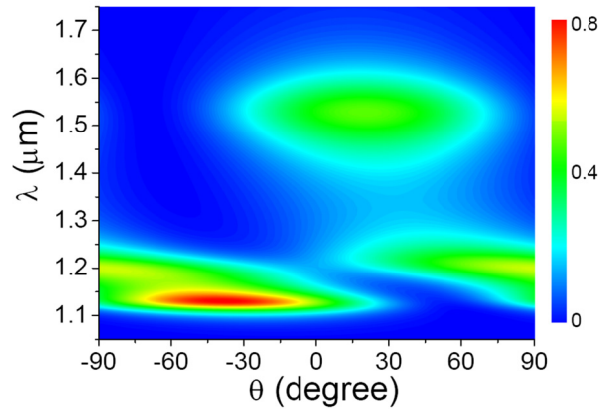

Fig. 4 Power transmission efficiency as a function of wavelength and polarization angle θ of the incident light (calculated with the GML).

## 5. Conclusions

In summary, a single-layer interference-type plasmonic polarizer which owns both interference effect and polarization properties has been proposed and investigated. The single-layer plasmonic polarizer can couple both the parallel and vertical components of the incident field to the transmission, thus breaking the classical law of Malus. A variety of abnormal polarization effects such as the asymmetric polarization-angle dependence, enhanced polarization filtering with wide polarization angle, etc. have been revealed. By rotating the plasmonic polarizer mechanically, a tunable and efficient polarization rotation from $0^\circ$ to $90^\circ$ can also be achieved. We suggest that the interference-type plasmonic polarizer is governed by the generalized law of Malus, originating from the superposition principle and establishing the relationship between the transmission and polarization angle. The theory also demonstrates that such a plasmonic polarizer cannot transform the natural light into a linear polarization with the efficiency larger than 50%. This is a limit for different types of the individual polarizers. The GML provides a basis for the further study of the interference-type plasmonic polarizers. For example, with the optimization of the metasurface structure and transmission coefficients as well as the phase difference, one may seek the broadband and highly efficient polarizers that operate with a wide polarization angle. The results presented in this work could also be useful for understanding and designing novel optical devices where the interference effect serves as an extra degree of freedom and plays a crucial role.

## Acknowledgements

This work was supported by the National Natural Science Foundation of China (Grant No. 11804157).